\documentclass{IEEEtran}
\usepackage[cmex10]{amsmath}
\usepackage{array}
\usepackage{mdwmath}
\usepackage{algorithm}
\usepackage{algorithmic}
\usepackage{enumerate}
\usepackage{amssymb}
\usepackage{mdwtab}
\usepackage{graphicx}
\usepackage{stfloats}
\usepackage{cite}
\usepackage{color}
\usepackage{bm}

\usepackage{multirow}
\usepackage{booktabs}
\usepackage[tight,footnotesize]{subfigure}

\begin{document}

\renewcommand{\baselinestretch}{1.00}

\title{Learning-Based Joint Antenna Selection and Precoding Design for Cell-Free MIMO Networks}

\author{Liangzhi Wang, \emph{Graduate Student Member, IEEE,}~Chen~Chen, \emph{Member, IEEE,}~Jie~Zhang, \emph{Senior Member, IEEE}, and~Carlo Fischione, \emph{Fellow, IEEE}
\thanks{Manuscript received xxx; revised xxx; accepted xxx. Date of publication xxx; date of current version xxx. The work of Liangzhi Wang and Jie Zhang was supported in part by the U.K. Engineering and Physical Sciences Research
Council under Grant 101086219. 
The work of Chen Chen and Carlo Fischione was supported in part by the KTH Digital Future research center. 
The review of this paper was coordinated by xxx. 

Liangzhi Wang is with the Department of Electronic and Electrical Engineering, University of Sheffield, Sheffield, S10 2TN, UK (e-mail: lwang85@sheffield.ac.uk).

Chen Chen and Carlo Fischione are with the School of Electrical Engineering and Computer Science, KTH Royal Institute of Technology, Stockholm, Sweden (e-mail: \{chch2, carlofi\}@kth.se).

 Jie Zhang is with the
Department of Electronic and Electrical Engineering, University of Sheffield,
Sheffield, S10 2TN, UK, and also with Ranplan Wireless Network Design
Ltd., Cambridge, CB23 3UY, UK (e-mail: jie.zhang@sheffield.ac.uk).
}
\thanks{Color versions of one or more of the figures in this paper are available online at http://ieeexplore.ieee.org.}
\thanks{Digital Object Identifier xxx}	
}
\markboth{IEEE}%
{Shell \MakeLowercase{\textit{et al.}}: Bare Demo of IEEEtran.cls for Journals}
\maketitle

\begin{abstract}
This paper considers a downlink cell-free  multiple-input multiple-output (MIMO) network in which multiple multi-antenna access points (APs) serve multiple users via coherent joint transmission. 
In order to reduce the energy consumption by radio frequency components, each AP selects a 
subset of antennas for downlink data transmission after estimating the channel state information (CSI).
We aim to maximize the sum spectral efficiency by jointly optimizing the antenna selection and precoding design.
To alleviate the fronthaul overhead and enable real-time network operation, we propose a distributed scalable machine learning algorithm. 
In particular, at each AP, we deploy a convolutional neural network  (CNN) for antenna selection and a graph neural network  (GNN) for precoding design. 
Different from conventional centralized solutions that require a large amount of CSI and signaling exchange among the APs, the proposed distributed machine learning algorithm takes only locally estimated CSI as input.
With well-trained learning models, it is shown that the proposed algorithm significantly outperforms the distributed baseline schemes and achieves a sum spectral efficiency comparable to its centralized counterpart.
\end{abstract}


\IEEEpeerreviewmaketitle

\section{Introduction}
\IEEEPARstart{M}{assive} multiple-input multiple-output (MIMO) has been one of the most promising technologies in the fifth generation (5G) and beyond wireless communication networks~\cite{wang2023road, parkvall2017nr}. 
A massive MIMO base station (BS) is equipped with a large number of antenna elements, which allows it to spatially multiplex multiple users over the same time-frequency resource. However, the conventional cellular massive MIMO architecture cannot achieve uniform coverage at high spectral efficiency; the cell-edge users suffer from long transmission distances and severe inter-cell interference~\cite{chen2022deployment}. To this end, a new post-cellular network infrastructure, termed as cell-free MIMO, was proposed in~\cite{7827017}. 
In cell-free MIMO networks, the access points (APs) are connected to a central processing unit (CPU) or an edge-cloud processor via high-speed fronthaul links~\cite{burr2018ultra}. All the APs cooperatively serve the users by coherent joint transmission, eliminating cell-boundaries and inter-cell interference. This enables an improvement in both spectral efficiency and energy efficiency~\cite{ngo2017total}. 
On the other hand, the number of radio frequency (RF) chains increases with the number of antennas, which leads to escalating hardware cost and energy consumption~\cite{heath2001antenna}. 
In this regard, antenna selection can be employed to reduce the energy consumption by activating only the antennas that contribute the most to the system throughput~\cite{lopez2022survey}.

Antenna selection has been extensively investigated in  massive MIMO systems~\cite{gao2015massive, makki2016genetic, guo2020statistical, wang2023antenna, zhu2022antenna}. An exhaustive search for the optimal subset of antennas is generally computationally prohibitive. In~\cite{gao2015massive}, the authors considered zero-forcing precoding and converted the antenna selection problem into a convex optimization problem by relaxing the integer constraints. In~\cite{makki2016genetic},  antenna selection schemes based on genetic algorithms (GAs) were developed, which were shown to achieve almost the same throughput as exhaustive search. 
GA-based antenna selection was also adopted in~\cite{marinello2020antenna} to maximize the energy efficiency of a extra-large scale MIMO system.
The authors in~\cite{guo2020statistical} proposed to apply stochastic gradient descent in joint antenna selection and user scheduling to maximize the energy efficiency of a single-cell MIMO system. In~\cite{wang2023antenna}, the authors investigated limited-resolution analog-to-digital converters and digital-to-analog converters, and developed two quantization-aware antenna selection algorithms for downlink and uplink transmissions, respectively.
Recently, antenna selection in a full-duplex cell-free network was addressed centrally by an elite preservation GA at the CPU~\cite{zhu2022antenna}. 
The antenna selection schemes proposed in these works require iterative optimization, resulting in unacceptable computational burden. 

An alternative solution is to develop low-complexity schemes utilizing the recent advances of machine learning~\cite{elijah2022intelligent, hellstrom2022wireless}. Deep neural networks (DNNs) are able to approximate the complex mapping relationships between the time-varying channel conditions and the desired system configurations. The computational complexity is shifted to the offline training phase, allowing real-time online implementation. In~\cite{guo2023robust, balevi2020massive}, deep learning was employed in channel estimation tasks of massive MIMO systems. In~\cite{Chen2023IRS, jiang2021learning}, learning-based solutions were developed to jointly configure the transmit precoding at the multi-antenna BS and the phase shifts at the intelligent reflecting surface. In the context of a cell-free MIMO network with single-antenna APs, the joint channel estimation and beamforming was addressed in~\cite{chen2024joint} by developing model-driven deep learning solution.  In~\cite{ghiasi2022energy}, the AP-user association was optimized by deep reinforcement learning (DRL) to maximize the uplink energy efficiency in a user-centric cell-free MIMO network assuming maximum ratio combining.
Deep learning was exploited in uplink and downlink power control of a cell-MIMO network
in~\cite{d2019uplink} and~\cite{Zaher2023}, respectively. In particular, distributed DNNs were developed in~\cite{Zaher2023} to approximate the desired power allocation coefficients using only local inputs  determined by large-scale fading coefficients.

Machine learning has also found applications  in antenna selection problems~\cite{joung2016machine, gecgel2019transmit, vu2021machine, elbir2019joint, chai2020reinforcement}. In the context of single-cell MIMO system, the authors in~\cite{joung2016machine} modelled antenna selection as a multiclass-classification problem, which was solved by $k$-nearest neighbors and support vector machine algorithms.  Aiming to minimize the bit error rate, the authors in~\cite{gecgel2019transmit} proposed decision tree and multi-layer perceptron (MLP)
based schemes for antenna selection, taking imperfect channel state information (CSI) as input. The proposed schemes were validated via practical measurements on MIMO test-beds. In~\cite{vu2021machine}, an MLP was applied to antenna selection, while the precoding was optimized based on an iterative optimization algorithm. In~\cite{elbir2019joint}, two convolutional neural networks (CNNs) were developed for antenna selection and precoding design, respectively. The CNNs were further quantized to be applicable to low-memory devices. In~\cite{chai2020reinforcement}, the authors considered a cell-free MIMO scenario using conjugate beamforming.
A centralized DRL-based antenna selection algorithm was developed taking the global CSI as input. The developed DRL-based algorithm relied on online learning that cannot perform real-time antenna selection.

To the best of our knowledge, the joint precoding and antenna selection in cell-free MIMO networks has not been investigated yet. 
The main challenges of cell-free MIMO network architecture are the intensive CSI and signaling exchange over fronthaul links and the computational burden of signal processing. 
To reduce the fronthaul overhead of information exchange and computational complexity of joint optimization, distributed and real-time solutions are preferred. 
The existing centralized antenna selection and precoding schemes are not applicable in cell-MIMO scenarios.
In this paper, we novelly propose a fully distributed machine learning-based algorithm for joint antenna selection and precoding design.
The main contributions of this work are summarized as follows:

\begin{itemize}
	\item First, we investigate a multi-user cell-free MIMO network and formulate a joint antenna selection and precoding design problem to maximize the downlink sum spectral efficiency, which accounts for realistic pilot-based channel estimation, subject to the transmission power budget at each AP.
	\item Second, we develop a distributed graph neural network (GNN)-based precoding algorithm, where each AP deploys a GNN, using only local CSI estimate of selected antennas as input. The GNNs are trained centrally at the CPU  to maximize the sum spectral efficiency utilizing global CSI estimates. The well-trained GNNs can adapt to dynamic channel conditions and an arbitrarily selected subset of antennas.
	\item Third, we generate training datasets for antenna selection using iterative search with the aid of the well-trained GNNs. We develop a distributed CNN for each AP to capture the mapping relationship between the local CSI estimate and the selected antenna subset. 
	\item Finally, we perform extensive simulations to demonstrate that the proposed fully distributed joint CNN-based antenna selection and GNN-based precoding algorithm achieves a sum spectral efficiency close to its centralized counterpart by a much lower computational complexity.
\end{itemize}

The rest of this paper is organized as follows. 
Section \ref{sec:system_model} introduces the system model and problem formulation. 
In Section \ref{sec:Precoding}, we describe the proposed distributed GNN-based algorithm in detail, which is used for precoding design.  
The proposed distributed CNN-based antenna selection algorithm is detailed in Section \ref{sec:CNN_AS}.
In Section \ref{sec:simulation}, simulation results are presented to compare different schemes in terms of sum spectral efficiency.
Finally, the major conclusions are provided in Section \ref{sec:conclusion}.

Notations: In this paper, italic letters, boldface lower-case letters, and boldface uppercase letters represent scalars, vectors, and matrices, respectively. $\mathbf{V}^{T}$, $\mathbf{V}^{H}$, and $\mathbf{V}^{*}$ denote the transpose, conjugate transpose,
and conjugate of a matrix $\mathbf{V}$, respectively.
$\mathbb{E}\{\cdot\}$ is
the statistical expectation and $\text{Tr}(\cdot)$ is the trace operation.
$||\cdot||_F$ represents the Frobenius norm.
$\mathcal{N_{\mathbb{C}}}\left(\textbf{0}, \textbf{R}\right)$ denotes the distribution of a multivariate circularly symmetric complex Gaussian variable with zero mean and covariance matrix $\textbf{R}$.  $\mathbb{C}^{M\times N}$ and $\mathbb{R}^{M\times N}$ represent the space of complex-valued and real-valued matrices, respectively. $[\mathbf{V}]_{i,j}$ denotes the $(i,j)$-th element of a matrix $\mathbf{V}$ and  
 $[\mathbf{v}]_{i}$ denotes the $i$-th element of a vector $\mathbf{v}$.

\section{System Model and Problem Formulation}
\label{sec:system_model}
\begin{figure}[!t]
\centerline{\includegraphics[width= 3.4in]{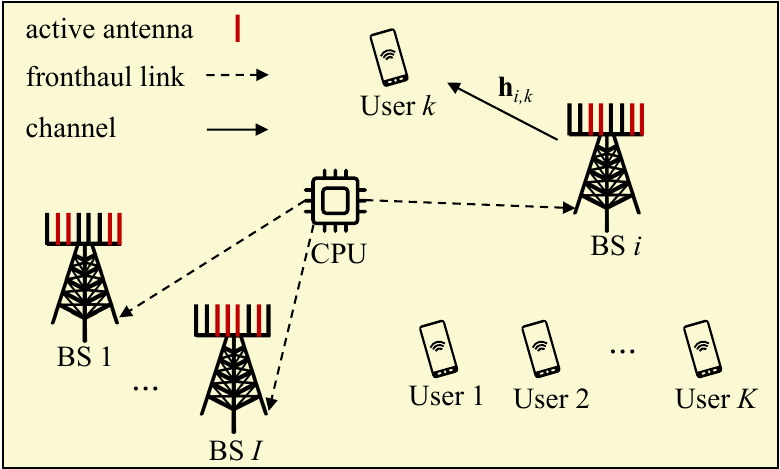}}
\caption{An illustration of the downlink cell-free MIMO network.}
\label{fig:system_model}
\end{figure}

As shown in Fig. \ref{fig:system_model}, we consider a multi-user downlink cell-free MIMO network, which consists of one CPU, $K$ single-antenna users and $I$ multi-antenna APs. Each AP is equipped with $N$ antennas.
The index sets of users and APs are represented by $\mathcal{K} = \left\{1,2,\dots,K\right\}$ and $\mathcal{I} = \left\{1,2,\dots,I\right\}$, respectively. 
  We assume that the AP antenna spacing is adequate such that the channels between the AP antennas and an arbitrary user are uncorrelated~\cite{bjornson2015optimal}. 
It is straightforward to extend the proposed machine learning algorithm to spatially correlated channels.
  The small-scale fading is assumed to follow Rayleigh distribution.  Accordingly, the channel from user $k$ to AP $i$ is denoted by $\textbf{h}_{i,k}\sim \mathcal{N_{\mathbb{C}}}\left(\textbf{0}, \beta_{i,k}\textbf{I}_{N}\right)$, where $\beta_{i,k}$ represents the large-scale path loss. 

The CPU and APs are connected by error-free front-haul links. Conventional centralized operation optimizes the precoding vectors at the CPU by collecting the global instantaneous CSI, which leads to a heavy front-haul overhead.
In this paper, we develop efficient distributed algorithms that enable the APs to select antenna subsets and design precoding vectors only according to their locally estimated CSI. In this way, no CSI is transmitted through the front-haul links but only uplink and downlink data.

\subsection{Channel Estimation}
\begin{figure}[!t]
\centerline{\includegraphics[width= 3.4in]{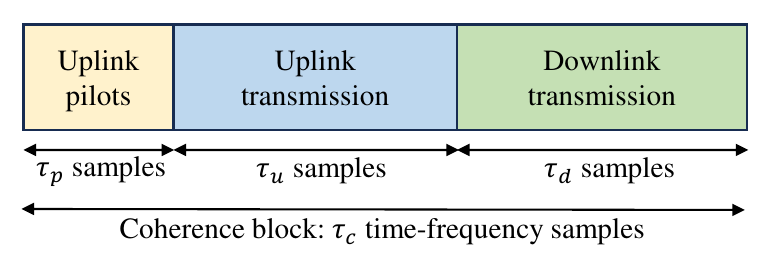}}
\caption{An illustration of the TDD protocol.}
\label{fig:TDD protocol}
\end{figure}

We consider the widely-used block fading channel model where the channels are static and frequency-flat in a coherence block with $\tau_{{c}}$ time-frequency samples.\footnote{Note that the size of the coherence block is assumed to be the same for
 all  the users. In practice, the size of the  coherence  block is determined by the worst-case propagation scenario that the network should support~\cite{bjornson2017massive, chen2024multi}.}
The time-division duplex (TDD) protocol is adopted for channel estimation and data transmission.\footnote{In this paper, we focus on low- to medium-mobility users.
Note that for wireless communications in high-mobility environments, it is required to consider the impact of channel aging and perform channel prediction.} 
As shown in Fig. \ref{fig:TDD protocol}, a coherence block is divided into three phases: $\tau_{{p}}$ samples for uplink pilot transmission, $\tau_{{u}}$ samples for uplink data transmission, and $\tau_{{d}}$ samples for downlink data transmission. We assign an uplink pilot signal $\bm{\phi}_{k}\in \mathbb{C}^{\tau_{p}\times 1}$ to user $k$, $\|\bm{\phi}_{k}\|^{2}= \tau_{{p}}, \forall k\in \mathcal{K}$. We also assume that $\tau_{p}\ge K$ and $\bm{\phi}_{k}^{H}\bm{\phi}_{k^{'}}=0, \forall k \ne k^{'}$.  During the uplink pilot transmission phase, the received signal $\textbf{Y}_{i}^{p}\in \mathbb{C}^{N\times \tau_{p}}$ at AP $i$ is given by
\begin{align}
\textbf{Y}_{i}^{p}=\sqrt{P_{\mathrm{ul}}}\sum_{k=1}^{K}\textbf{h}_{i,k}\bm{\phi}_{k}^{H} + \textbf{N}_{i}^{p},
\end{align}
where $P_{\mathrm{ul}}$ is the uplink pilot transmit power and $\textbf{N}_{i}^{p}\in \mathbb{C}^{N\times \tau_{p}}$ is the complex Gaussian noise matrix with i.i.d. entries $\mathcal{N_{\mathbb{C}}}\left(0, \delta^{2}\right)$. Then AP $i$ correlates $\textbf{Y}_{i}^{p}$ with $\bm{\phi}_{k}$ and obtains
\begin{align}
\textbf{y}_{i,k}^{p}
&=\sqrt{P_{\mathrm{ul}}}\textbf{h}_{i,k}\bm{\phi}_{k}^{H}\bm{\phi}_{k} + \sqrt{P_{\mathrm{ul}}}\sum_{k^{'}\in\mathcal{K}, k^{'}\ne k}\textbf{h}_{i,k^{'}}\bm{\phi}_{k^{'}}^{H}\bm{\phi}_{k} + \textbf{N}_{i}^{p}\bm{\phi}_{k} \nonumber \\
&=\sqrt{P_{\mathrm{ul}}}\tau_{{p}}\textbf{h}_{i,k}
+ \textbf{N}_{i}^{p}\bm{\phi}_{k},
\end{align}
where $\textbf{N}_{i}^{p}\bm{\phi}_{k} \sim \mathcal{N_{\mathbb{C}}}\left(\textbf{0}, \delta^{2}\tau_{{p}}\textbf{I}_{N}\right)$.
The linear minimum-mean-square-error (LMMSE)
estimate of $\textbf{h}_{i,k}$ is given by
\begin{align}
\widehat{\textbf{h}}_{i,k}=\sqrt{P_{\mathrm{ul}}}\tau_{{p}}\beta_{i,k}\textbf{Z}_{i,k}^{-1}\textbf{y}_{i,k}^{p}
=\frac{\sqrt{P_{\mathrm{ul}}}\beta_{i,k}\textbf{y}_{i,k}^{p}}{{P_{\mathrm{ul}}}\tau_{{p}}\beta_{i,k} + \delta^{2}},
\end{align}
where $\textstyle \textbf{Z}_{i,k}=\mathbb{E}\{\textbf{y}_{i,k}^{p}(\textbf{y}_{i,k}^{p})^{H}\}=P_{\mathrm{ul}}\tau_{{p}}^{2}\beta_{i,k}\textbf{I}_{N}+\delta^{2}\tau_{{p}}\textbf{I}_{N}$. The channel estimation error of $\textbf{h}_{i,k}$ is given by $\widetilde{\textbf{h}}_{i,k}= \textbf{h}_{i,k}- \widehat{\textbf{h}}_{i,k}$ with correlation matrix
\begin{align}
\textbf{C}_{i,k}
&=
\beta_{i,k}\textbf{I}_{N}- P_{\mathrm{ul}}\tau_{{p}}^{2}\beta_{i,k}^{2}\textbf{Z}_{i,k}^{-1} \nonumber \\
&=
\left(\beta_{i,k} - \frac{P_{\mathrm{ul}}\tau_{{p}}\beta_{i,k}^{2}}{{P_{\mathrm{ul}}}\tau_{{p}}\beta_{i,k} + \delta^{2}}\right)\textbf{I}_{N}.
\end{align}

\subsection{Antenna Selection}
In each coherence
block, we select a  $M$-out-of-$N$ subset of antennas as active antennas for data transmission at each AP. In this paper, we assume the same number of active antennas per AP for simplicity. The extension to scenarios where different APs have different numbers of active antennas is straightforward.
Denote the selected subset of antennas in AP $i$ by $\mathcal{A}_{i}=\{a_{i_{1}},a_{i_{2}},\dots,a_{i_{M}}\}$, where $a_{i_m}\in \{1,2,\dots,N\}$. 
The number of all the possible antenna subsets at AP $i$ is $\textstyle C_{i}=\tbinom{N}{M}$.
Then, the channel from user $k$ to the active antennas at AP $i$ is given by
\begin{equation}
\textbf{h}_{i,k,\mathcal{A}_{i}} = \left[h_{i,k,a_{i_{1}}},h_{i,k,a_{i_{2}}},\dots,h_{i,k,a_{i_{M}}}\right]^{T}\in \mathbb{C}^{M\times 1},
\end{equation}
where $h_{i,k,a_{i_{m}}}$ denotes the channel between antenna $a_{i_{m}}$ at AP $i$ and user $k$. Let $\mathcal{A}=\{\mathcal{A}_{1}, \mathcal{A}_{2},\dots, \mathcal{A}_{I}\}$ represent the global antenna subset in the cell-free MIMO network.
The number of all the possible global antenna selection strategies is computed by $\textstyle C=\prod_{i=1}^{I}C_{i}=\tbinom{N}{M}^{I}$.

In a cell-free MIMO network, each of the users is served by all the multi-antenna APs simultaneously over the same time-frequency resource.
Thus, the received signal at user $k$ can be expressed as
\begin{align}
y_{k} 
=& \sum_{\vphantom{k^{'}\in \mathcal{K}}i \in \mathcal{I}} 
\sum_{\vphantom{k^{'}\in \mathcal{K}}k^{'} \in \mathcal{K}} 
\textbf{h}_{i,k,\mathcal{A}_{i}}^{H}\textbf{w}_{i,k^{'}}{s}_{k^{'}} 
+ n_k
\nonumber \\
 =& \!\underbrace{\vphantom{\sum_{\vphantom{k^{'}\in \mathcal{K}}i \in \mathcal{I}}}\sum_{\vphantom{k^{'}\in \mathcal{K}}i \in \mathcal{I}}\widehat{\textbf{h}}_{i,k,\mathcal{A}_{i}}^{H}\textbf{w}_{i,k}{s}_{k}}_{\text{the desired signal}} 
 + \underbrace{\vphantom{\sum_{\vphantom{k^{'}\in \mathcal{K}}i \in \mathcal{I}}}\sum_{\vphantom{k^{'}\in \mathcal{K}}i \in \mathcal{I}} \sum_{\vphantom{k^{'}\in \mathcal{K}}k^{'}\in \mathcal{K},k^{'}\neq k}\widehat{\textbf{h}}_{i,k,\mathcal{A}_{i}}^{H}\textbf{w}_{i,k^{'}}{s}_{k^{'}}}_{\text{the interference signal}} \nonumber \\
 &+ \underbrace{\sum_{\vphantom{k^{'}\in \mathcal{K}}i \in \mathcal{I}} \sum_{\vphantom{k^{'}\in \mathcal{K}}k^{'}\in \mathcal{K}}\widetilde{\textbf{h}}_{i,k,\mathcal{A}_{i}}^{H}\textbf{w}_{i,k^{'}}{s}_{k^{'}}}_{\text{the channel estimation errors}}
 +
\underbrace{\vphantom{\sum_{\vphantom{k^{'}\in \mathcal{K}}i \in \mathcal{I}}}n_{k},}_{\text{the noise}}
\end{align}
where $n_{k} \sim \mathcal{N_{\mathbb{C}}}\left(0, \delta^{2}\right)$ denotes the additive white Gaussian noise at user $k$, $\textbf{w}_{i,k}\in \mathbb{C}^{M\times1} $ represents the beamforming vector from AP $i$ to user $k$, and $s_{k}$ is the required data symbol of user $k$.
$\widehat{\textbf{h}}_{i,k,\mathcal{A}_{i}}$ and $\widetilde{\textbf{h}}_{i,k,\mathcal{A}_{i}}$ are the LMMSE estimate and estimation error of ${\textbf{h}}_{i,k,\mathcal{A}_{i}}$, respectively.

Due to the channel estimation error, it is difficult to obtain an accurate expression of the downlink spectral
efficiency. Following the existing works~\cite{ammar2021downlink, jose2011pilot}, we treat the channel estimation error as Gaussian noise and
adopt a lower bound of the downlink spectral efficiency as follows
\begin{equation}
\label{Rk}
R_{k}\left(\mathcal{A},\textbf{w}_{k}\right) = \frac{\tau_{d}}{\tau_{c}}\mathrm{log}_{2}\left( 1 + \text{SINR}_{k}\left(\mathcal{A},\textbf{w}_{k}\right)\right),
\end{equation}
where the instantaneous effective signal-to-interference-and- noise ratio (SINR) is given by
\begin{align}
&\text{SINR}_{k}\left(\mathcal{A},\textbf{w}_{k}\right) \nonumber \\
&\ = 
\frac{{\left| \widehat{\textbf{h}}_{k,\mathcal{A}}^{H} \textbf{w}_{k}\right|}^{2}}{\sum\limits_{k^{'}\in \mathcal{K},k^{'}\neq k} {\left| \widehat{\textbf{h}}_{k,\mathcal{A}}^{H} \textbf{w}_{k^{'}}\right|}^{2} 
+ 
\sum\limits_{k^{'}\in \mathcal{K}} \textbf{w}_{k^{'}}^{H}\textbf{C}_{k,\mathcal{A}}\textbf{w}_{k^{'}}
+
\delta^{2}},
\end{align}
where $\widehat{\textbf{h}}_{k,\mathcal{A}} \!=\! \left[\widehat{\textbf{h}}_{1,k,\mathcal{A}_{1}}^{T},\widehat{\textbf{h}}_{2,k,\mathcal{A}_{2}}^{T},\dots,\widehat{\textbf{h}}_{I,k,\mathcal{A}_{I}}^{T}\right]^{\!T\!}$, 
$\widetilde{\textbf{h}}_{k,\mathcal{A}} \!\!=\!\! \left[\widetilde{\textbf{h}}_{1,k,\mathcal{A}_{1}}^{T},\widetilde{\textbf{h}}_{2,k,\mathcal{A}_{2}}^{T},\dots,\widetilde{\textbf{h}}_{I,k,\mathcal{A}_{I}}^{T}\right]^{\!T\!}$,  $\textbf{w}_{k} \!\!=\!\! \left[\textbf{w}_{1,k}^{T},\textbf{w}_{2,k}^{T},\dots,\textbf{w}_{I,k}^{T}\right]^{\!T\!}$, and $\textbf{C}_{k,\mathcal{A}}=\text{diag}\left(\textbf{C}_{1,k,\mathcal{A}_{1}}, \textbf{C}_{2,k,\mathcal{A}_{2}}, \dots, \textbf{C}_{I,k,\mathcal{A}_{I}}\right)$ with 
\begin{align}
\textbf{C}_{i,k,\mathcal{A}_{i}}&= \mathbb{E}\left\{\widetilde{\textbf{h}}_{i,k,\mathcal{A}_{i}}\left(\widetilde{\textbf{h}}_{i,k,\mathcal{A}_{i}}\right)^{H}\right\} \nonumber \\
&= \left(\beta_{i,k} - \frac{P_{\mathrm{ul}}\tau_{{p}}\beta_{i,k}^{2}}{{P_{\mathrm{ul}}}\tau_{{p}}\beta_{i,k} + \delta^{2}}\right)\textbf{I}_{M}.
\end{align}

\subsection{Problem Formulation}
In this paper, we aim to maximize the sum spectral efficiency by jointly optimizing the antenna selection strategy $\mathcal{A}$ and the precoding vectors $\left\{\textbf{w}_{k}|\forall k \in \mathcal{K}\right\}$ at the APs. 
Mathematically, the optimization problem can be formulated as follows
\begin{align}
\mathop{\max_{\mathcal{A}, \left\{\textbf{w}_{k}\right\}}}~& \sum_{k\in \mathcal{K}}R_{k}\left(\mathcal{A},\textbf{w}_{k}\right)  \label{P}   \\
      s.t.~          
      & \sum_{k\in\mathcal{K}}\text{Tr} \left( \textbf{w}_{i,k} \textbf{w}_{i,k}^{H} \right) \le P_{\text{max}}, \forall i\in \mathcal{I}, \tag{\ref{P}{a}}
\end{align}
where $P_{\rm max}$ is the maximum transmit power per AP. The joint optimization problem is a mixed integer non-convex optimization problem. Note that the precoding vectors are closely related to active antennas. Specifically, precoding vectors are designed for a certain, predetermined active antenna subset. 

In this paper, we propose to solve the optimization problem using distributed machine learning algorithms. More specifically, we first propose a distributed GNN-based precoding scheme 
to optimize the precoding vectors for an arbitrary given antenna selection strategy, as detailed in Section \ref{sec:Precoding}. Subsequently, in Section \ref{sec:CNN_AS}, we apply the well-trained GNNs to generate datasets for training CNN models that can be used to predict the optimal antenna selection strategy.

\section{Proposed Distributed GNN-Based Precoding}
\label{sec:Precoding}
Conventional centralized precoding approaches such as centralized MMSE and centralized iterative optimization require the CPU to collect the global CSI and determine the precoding vectors for all the APs, and then transmit the precoding vectors to each AP through the front-haul links. 
This process requires a large amount of information exchange, leading to a heavy fronthaul overhead. 
On the other hand, the existing distributed precoding schemes such as Maximum ratio transmission (MRT) and distributed MMSE suffer from the limited interference cancellation ability of a single AP.
To this end, in this section, we develop a fully distributed precoding algorithm based on GNNs that can achieve similar sum spectral efficiency as centralized approaches. 
The GNNs are trained offline in a centralized manner at the CPU and downloaded by the APs for online use in a distributed manner.
Each AP is able to design its precoding vectors using the well-trained GNN only according to its locally estimated CSI.

\subsection{Graph Construction}
GNN is a class of neural network models for processing graph-structured data. A graph $\mathcal{G}$ typically consists of  node set $\mathcal{N}$ and edge set $\mathcal{E}$. Each node has its own attributes that are used to store feature information such as node identity. Edges record information that describes the relationship between nodes. Edges can be directed, which means that the information transfer between nodes is unidirectional. When the information transfer between nodes is bidirectional, the edges are undirected. The connected relation between nodes can be easily represented by the adjacency matrix $\textbf{A}$ with main diagonal elements being zero. When node $k$ is connected to node $k^{'}$ by an undirected edge, $[\textbf{A}]_{k,k^{'}} = [\textbf{A}]_{k^{'},k} = 1$; when node $k$ and $k^{'}$ are not connected, $[\textbf{A}]_{k,k^{'}} = [\textbf{A}]_{k^{'},k} = 0$. 

\begin{figure}[!t]
\centerline{\includegraphics[width= 3.4in]{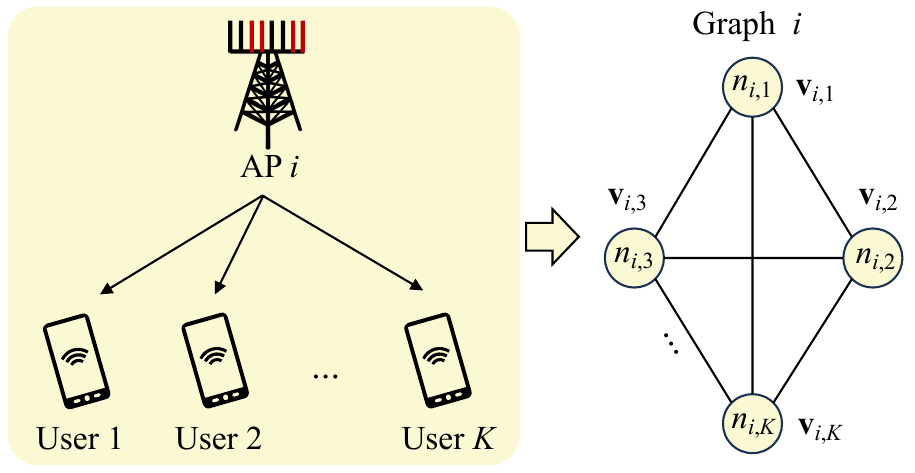}}
\caption{An illustration of Graph $i$. It is a fully connected undirected graph, where node $n_{i,k}$ represents user $k$, and $\textbf{v}_{i,k}$ is its feature vector.}
\label{fig:Graph i}
\end{figure}

This paper aims to deploy a GNN at each AP for designing the precoding vectors. As shown in Fig. \ref{fig:Graph i}, graph $i$, $\forall i\in \mathcal{I}$, is constructed from the subnetwork consisting of AP $i$ and all the users. 
Let $\mathcal{G}_{i} = (\mathcal{N}_{i}, \mathcal{E}_{i})$ denote graph $i$, where $\mathcal{N}_{i} = \left\{{n}_{i,1}, \dots, {n}_{i,K}\right\}$ represents the set consisting of all user nodes and  $\mathcal{E}_{i}$ is the set of edges.
 Specifically, graph $i$ is constructed as a fully connected undirected graph, i.e., there is an undirected connectivity between any two nodes. Hence, the adjacency matrix of graph $i$ can be described as: $[\textbf{A}_{i}]_{k,k^{'}} = [\textbf{A}_{i}]_{k^{'},k} = 1, \forall k \neq k^{'} $.  
 Denote
$\mathcal{V}_{i} = \left\{\textbf{v}_{i,1}, \dots, \textbf{v}_{i,K}\right\}$ as the set consisting of the feature vectors, where $\textbf{v}_{i,k}$ is the feature vector used to characterize ${n}_{i,k}$.

We adopt spatial graph convolutional networks to update the feature vectors~\cite{gilmer2017neural}. 
To sufficiently learn the inter-node relationships, each layer of the GNN performs convolution operations which amalgamate information from object nodes and their neighboring nodes to update the feature vectors. The feature vector update will be elaborated in the next subsection.
We train the GNNs using unsupervised learning, bypassing the time-consuming step of dataset generation. Compared to other deep learning networks, GNNs can better  handle inter-user interference in the precoding design. In Section \ref{sec:GNN_Setup}, we will show  that our proposed  GNN-based precoding design is superior to CNN- and MLP-based solutions in terms of sum spectral efficiency.

\subsection{Structure of the Proposed GNNs}
\begin{figure}[!t]
%\vspace{1mm}
\centerline{\includegraphics[width= 3.4in]{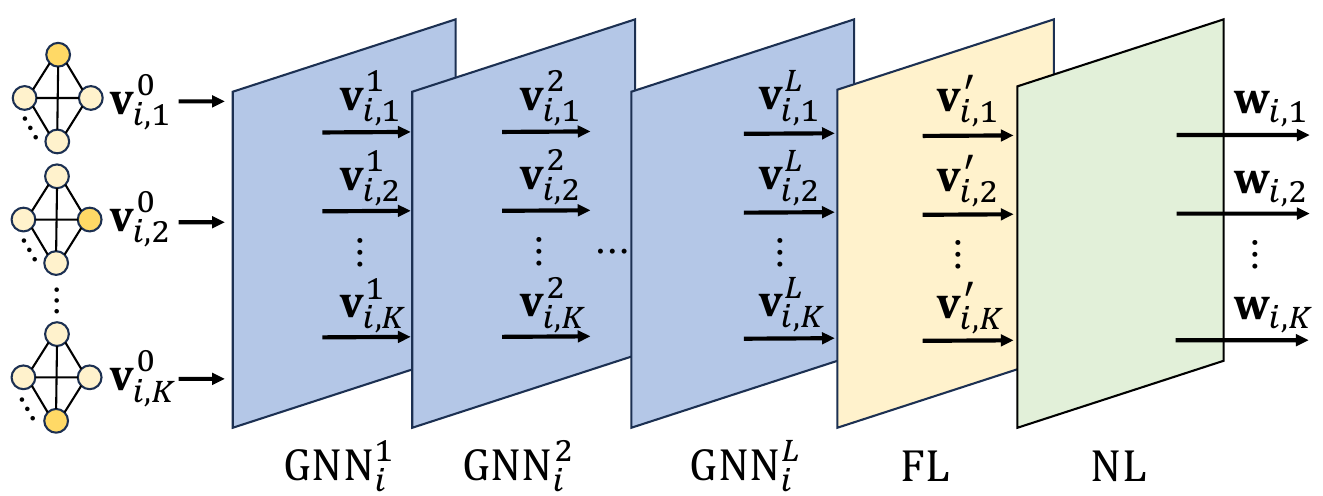}}
\caption{The structure of GNN $i$.}
\label{fig:GNN i}
\end{figure}

The structure of GNN $i, \forall i\in \mathcal{I}$, is shown in Fig. \ref{fig:GNN i}. In this subsection, we detail how the proposed GNN model extracts the key information from the locally estimated CSI and then designs the precoding vectors. Overall, GNN $i$ contains $L$ graph convolutional layers, denoted by $\left\{ \text{GNN}_{i}^{1}, \text{GNN}_{i}^{2}, \dots \text{GNN}_{i}^{L} \right\}$, one feed-forward layer (FL) and one normalization layer (NL). The output of each layer is used as the input of the next layer. The index set of the graph convolutional layers per GNN is denoted by $\mathcal{L}=\{1,2,\dots,L\}$.

\subsubsection{Initial Input} 
The initial input, i.e., the initial feature vector of $n_{i,k}$, is the estimated CSI between AP $i$ and the user $k$,
which is represented by
\begin{align}
\textbf{v}_{i,k}^{0} = \left[\Re\left\{\widehat{\textbf{h}}_{i,k,\mathcal{A}_{i}}^{T}\right\}, \Im\left\{\widehat{\textbf{h}}_{i,k,\mathcal{A}_{i}}^{T}\right\}\right]^{T}\in \mathbb{R}^{2M\times 1},
\end{align}
where $\textstyle \Re\{\widehat{\textbf{h}}_{i,k,\mathcal{A}_{i}}^{T}\}$ and $\Im\{\widehat{\textbf{h}}_{i,k,\mathcal{A}_{i}}^{T}\}$ represent the real and imaginary parts of $\widehat{\textbf{h}}_{i,k,\mathcal{A}_{i}}$, respectively. 
This is consistent with the requirement of neural networks for real-valued inputs. Note that the perfect CSI is not available at the APs due to the channel estimation errors.

\subsubsection{Graph Convolution Layer}
\begin{figure}[!t]
\centerline{\includegraphics[width= 2.3in]{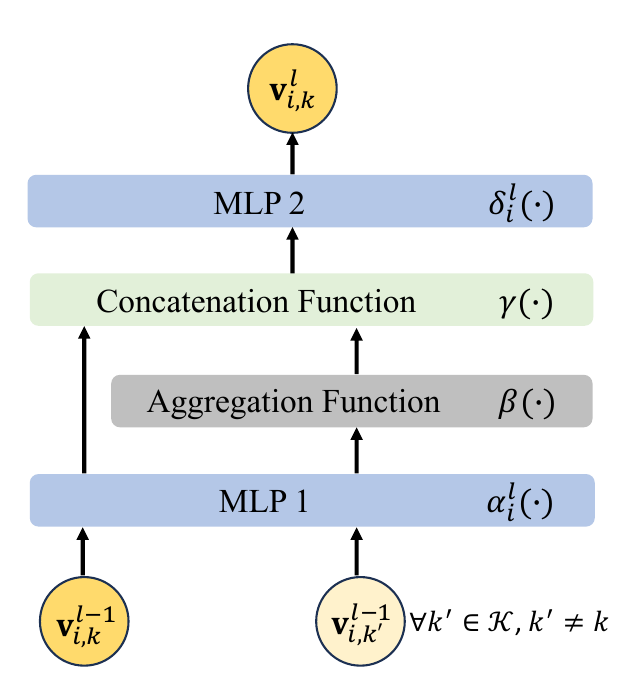}}
\caption{An illustration of graph convolutional layer $\text{GNN}_{i}^{l}$.}
\label{fig:graph convolution}
\end{figure}

The graph convolutional layers serve as fundamental components of GNN, responsible for updating the node feature vectors. Taking $\text{GNN}_{i}^{l}$ as an example, the process of updating the feature vector $\textbf{v}_{i,k}$ of $n_{i,k}$ is shown in Fig. \ref{fig:graph convolution}. The input of $\text{GNN}_{i}^{l}$ first goes through an MLP, referred to as $\alpha_{i}^{l}(\cdot)$, which maps the previous feature vectors into a high-dimensional space, thereby capturing additional network topology information. This step can be formulated as follows
\begin{align} 
\textbf{f}_{i,k}^{l} &= \alpha_{i}^{l}\left(\textbf{v}_{i,k}^{l-1}\right),  \\
\textbf{f}_{i,k^{'}}^{l} &= \alpha_{i}^{l}\left(\textbf{v}_{i,k^{'}}^{l-1}\right), \forall n_{i,k^{'}}\in \mathcal{C}(n_{i,k}),
\end{align}
where $\mathcal{C}(n_{i,k})=\left\{n_{i,k^{'}} | \forall k^{'}\in \mathcal{K}, k^{'}\ne k\right\}$ denotes the set of all neighbouring nodes of $n_{i,k}$. 
We adopt element-wise max function $\beta(\cdot)$ to aggregate the outputs of neighbouring nodes, which is expressed as
\begin{align} 
\textbf{g}_{i,k}^{l} &=  \beta\left(\left\{\textbf{f}_{i,k^{'}}^{l} |  \forall n_{i,k^{'}}\in \mathcal{C}(n_{i,k})\right\}\right).
\end{align}
Then we concatenate the aggregated vector $\textbf{g}_{i,k}^{l}$ with  $\textbf{f}_{i,k}^{l}$ and obtain the concatenated vector as follows
\begin{align} 
\textbf{c}_{i,k}^{l} &= \gamma\left(\textbf{g}_{i,k}^{l}, \textbf{f}_{i,k}^{l}\right),
\end{align}
where $\gamma(\cdot)$ denotes the concatenation function. 
Finally, an MLP, denoted by $\delta_{i}^{l}(\cdot)$, is employed to further extract information from $\textbf{c}_{i,k}^{l}$ and output the new feature vector of $n_{i,k}$ at the $l$-th layer as follows
\begin{align} \label{eq:Phiin}
\textbf{v}_{i,k}^{l} &= \delta_{i}^{l}\left(\textbf{c}_{i,k}^{l}\right).
\end{align}
It's worth noting that $\alpha_{i}^{l} (\cdot)$ and $\delta_{i}^{l} (\cdot)$ have the same hyper-parameters at all the user nodes, respectively. This reduces the number of hyper-parameters and enables faster convergence of the neural network's training. 

\subsubsection{Output}
After $L$ graph convolutional layers, the feature vectors of user nodes have adequate information to design the corresponding precoding vectors. The feature vector of $n_{i,k}$ at the $L$-th layer, $\textbf{v}_{i,k}^{L}$, is followed by a FL layer with $2M$ units that outputs a real-valued vector $\textbf{v}_{i,k}^{'}\in \mathbb{R}^{2M\times1}$. We further convert $\textbf{v}_{i,k}^{'}$ to a complex-valued vector, which is given by
\begin{align}
\textbf{w}_{i,k}^{'}= \left[\textbf{v}_{i,k}^{'}\right]_{1:M} + j\left[\textbf{v}_{i,k}^{'}\right]_{M+1:2M}.
\end{align}
Recall that each AP has a maximum transmit power $P_{\mathrm{max}}$. An NL is adopted to obtain the precoding matrix at AP $i$, $\textbf{W}_{i}$, as follows
\begin{align}
\textbf{W}_{i}^{'} &= \left[\textbf{w}_{i,1}^{'}, \textbf{w}_{i,2}^{'}, \dots, \textbf{w}_{i,K}^{'}\right] \in \mathbb{R}^{M\times K}, \\
\textbf{W}_{i} &= \sqrt{P_{\mathrm{max}}}\frac{\textbf{W}_{i}^{'}}{\|\textbf{W}_{i}^{'}\|_F}.
\end{align}
Accordingly, the precoding vector of user $k$ at AP $i$ is given by $\textbf{w}_{i,k} = \left[\textbf{W}_{i}\right]_{:,k}$.

\begin{figure}[!t]
\centerline{\includegraphics[width= 3.4in]{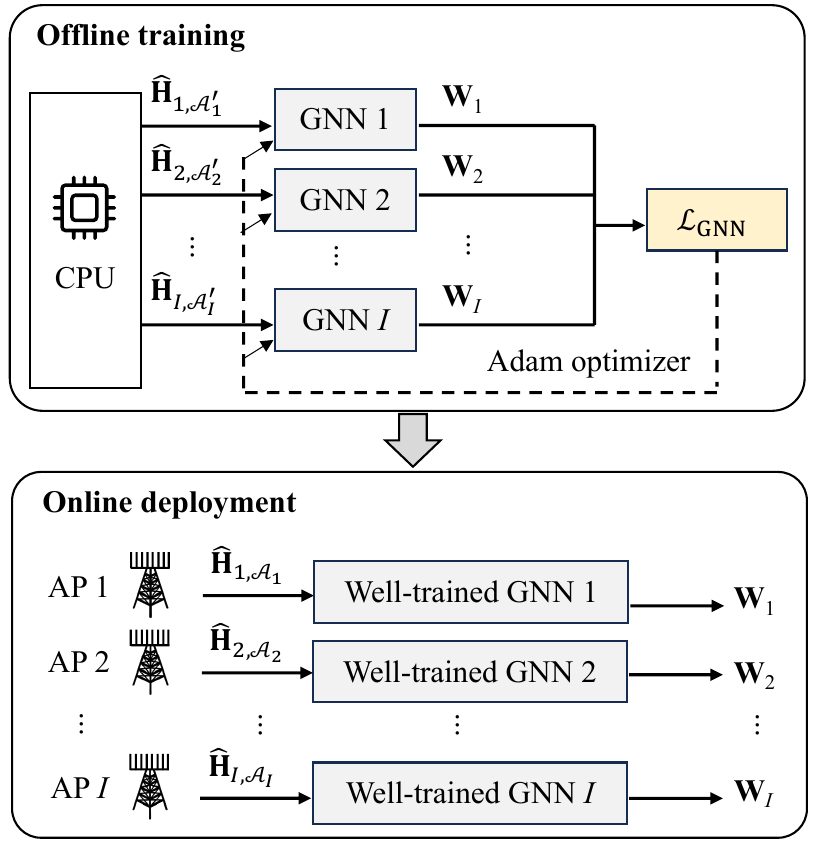}}
\caption{Offline training and online deployment of the proposed distributed GNN-based algorithm.}
\label{fig:offline training online deployment1}
\end{figure}

\subsection{Offline Training and Online Deployment}
The proposed distributed GNN-based precoding algorithm consists of two phases: offline training and online deployment, as shown in Fig. \ref{fig:offline training online deployment1}, where $\widehat{\textbf{H}}_{i,\mathcal{A}_{i}^{'}}=[\widehat{\textbf{h}}_{i,1,\mathcal{A}_{i}^{'}}, \widehat{\textbf{h}}_{i,2,\mathcal{A}_{i}^{'}}, \dots, \widehat{\textbf{h}}_{i,K,\mathcal{A}_{i}^{'}}]\in \mathbb{C}^{M\times K}$ is the estimated channel matrix between AP $i$ and all the users under the randomly selected antenna subset $\mathcal{A}_{i}^{'}$. The training phase occurs in the CPU in a centralized manner. 

We adopt unsupervised learning to train the GNNs and directly use the opposite of the sum spectral efficiency of all the users in the cell-free MIMO network as the loss function, which is given by
\begin{align}
\mathcal{L}_{\mathrm{GNN}} &= -\frac{1}{S}\sum_{s=1}^{S}\sum_{k\in\mathcal{K}} R_{k}(\mathcal{A}_{s}^{'},\textbf{w}_{k,s}),
\end{align} 
where $S$ is the number of training samples, $\mathcal{A}_{s}^{'}$ is the global antenna subset for the $s$-th training sample following the random antenna selection strategy, and $\textbf{w}_{k,s}$ is the precoding vector from all APs to user $k$ determined by the GNNs for the $s$-th training sample. The Adam optimizer is  adopted to minimize the loss function, i.e., maximize the sum spectral efficiency, and update the weight vectors of the GNNs. Since a large number of randomly selected antenna subsets are used in the training phase, the GNNs are capable of addressing an arbitrary given antenna selection strategy.

The online deployment phase is fully distributed, as shown in Fig. \ref{fig:offline training online deployment1}. Each AP initially acquires the pre-trained GNN models' weight parameters from the CPU. Subsequently, AP $i$ employs its weight parameters to autonomously determine local precoding vectors based on a given active subset of antennas $\mathcal{A}_{i}$ and its own locally estimated CSI $\widehat{\textbf{H}}_{i,\mathcal{A}_{i}}$.

\section{Proposed Distributed CNN-based Antenna Selection}
\label{sec:CNN_AS}
   Recall that the number of the possible antenna selection strategies at all the APs is $C=\prod_{i=1}^{I}C_{i}=\tbinom{N}{M}^{I}$, and thus antenna selection is a high-dimensional combinatorial optimization problem. It is difficult to get the solution within the channel coherence time using conventional schemes. 
   Moreover, finding the optimal antenna selection strategy requires global CSI, which leads to a heavy fronthaul overhead. To address these challenges, we exploit advanced machine learning algorithms to learn the mapping relationship between the channel observations and the optimal antenna subsets. More specifically, we develop a distributed CNN-based antenna selection algorithm, where each AP uses a CNN to select a $M$-out-of-$N$ subset of antennas only according to its locally estimated CSI.

    \begin{algorithm} [!t]
    \label{algorithm1}
    \renewcommand{\algorithmicrequire}{\textbf{Input:}}
    	\renewcommand{\algorithmicensure}{\textbf{Output:}}
    	\caption{Antenna selection dataset generation}
    	\begin{algorithmic}[1]
    	\REQUIRE $P_{\mathrm{max}}, I, N, M, K$ \ \% System parameters
     \FOR{$t = 1:T$}
            \STATE Generate channels $\textbf{H} = [\textbf{H}_{1}, \textbf{H}_{2}, ..., \textbf{H}_{I}]$
            \STATE Generate estimated channels $\widehat{\textbf{H}} = [\widehat{\textbf{H}}_{1}, \widehat{\textbf{H}}_{2}, ..., \widehat{\textbf{H}}_{I}]$
            \STATE $\textbf{X}_{i}^{t}=\left[\Re\left\{\widehat{\textbf{H}}_{i}\right\}; \Im\left\{\widehat{\textbf{H}}_{i}\right\}\right]$ \ \% Update the feature of  training sample $t$ in the antenna selection dataset of CNN $i$
            \STATE Initialize $\mathcal{A}_{i}, \forall i\in \mathcal{I}$, as the antenna subset with index $1$ $\left(\mathcal{A}_{i}=\{1,2,\dots,M\}\right)$
             \STATE Initialize $\textbf{W}_{i}, \forall i \in \mathcal{I}$, using the well-trained GNNs
     \STATE $R_{\text{max}}$ $= 0$ \ \% Initialize the maximum sum spectral efficiency
                \FOR{$i = 1:I$} 
    		\FOR{$j = 1:\tbinom{N}{M}$}
                \STATE Update $\mathcal{A}_{i}$ as the antenna subset with index $j$ 
      \STATE Update $\textbf{W}_{i}$ using the well-trained GNN $i$ 
      \STATE Update  $\sum_{k\in \mathcal{K}}R_{k}\left(\mathcal{A},\textbf{w}_{k}\right)$ according to (\ref{Rk})
      \IF{$\sum_{k\in \mathcal{K}}R_{k}\left(\mathcal{A},\textbf{w}_{k}\right)$ $>$ $R_{\text{max}}$} 
      \STATE $R_{\text{max}}=$ $\sum_{k\in \mathcal{K}}R_{k}\left(\mathcal{A},\textbf{w}_{k}\right)$
      \STATE $L_{i}^{t} = \mathcal{A}_{i}$ \ \% Update the label of  training sample $t$ in the antenna selection dataset of CNN $i$
    		\ENDIF
                \ENDFOR
    		\ENDFOR
      \ENDFOR
            \ENSURE  $\left\{\textbf{X}_{i}^{t}, L_{i}^{t}\right\}, i=1,2,\dots,I, t=1,2,\dots,T$
    	\end{algorithmic}  
    \end{algorithm}

\subsection{Dataset Generation}
\label{sec:dataset}
We deploy CNN $i$ at AP $i$ to select the optimal antenna subset $\mathcal{A}_{i}$. Since neural networks only support real-valued inputs,
the input of CNN $i$ is $\textbf{X}_{i}=[\Re\{\widehat{\textbf{H}}_{i}\}; \Im\{\widehat{\textbf{H}}_{i}\}]\in \mathbb{R}^{2N\times K}$,  where $\widehat{\textbf{H}}_{i}=[\widehat{\textbf{h}}_{i,1}, \widehat{\textbf{h}}_{i,2}, \dots, \widehat{\textbf{h}}_{i,K}]\in \mathbb{C}^{N\times K}$ is the estimated channel matrix between AP $i$ and the users. In this paper, antenna selection is modeled as a classification problem, where CNN $i$ selects an antenna subset from $C_{i}=\tbinom{N}{M}$ possible antenna selection strategies.
The output of CNN $i$ is the index of the selected antenna subset, denoted by $L_{i}$. 
Accordingly, training sample $t$ in the antenna selection dataset of CNN $i$ consists of two parts: feature, $\textbf{X}_{i}^{t}$, and label, $L_{i}^{t}$, where $\textbf{X}_{i}^{t}$ is the real-valued input matrix at sample $t$ and $L_{i}^{t}$ is the index of the optimal antenna subset at AP $i$ that maximizes the sum spectral efficiency.

 The generation of antenna selection dataset is performed at the CPU as the calculation of the sum spectral efficiency requires global CSI.
The exhaustive search algorithm iteratively evaluates all the possible  antenna selection strategies and chooses the best one that maximizes the sum spectral efficiency. The number of searches is $C=\prod_{i=1}^{I}C_{i}=\tbinom{N}{M}^{I}$. Hence, the exhaustive search algorithm is not tractable for large numbers of antennas and APs. To reduce the computational complexity, we propose  a new iterative search (IS) algorithm that performs iterative searches segment by segment. Segment $i$ is constructed by AP $i$ and the users. In segment $i$, AP $i$ iteratively searches $C_{i}=\tbinom{N}{M}$ possible antenna subsets and selects the best one that maximizes the sum spectral efficiency in (\ref{P}).
More specifically, the well-trained GNN $i$ is applied to obtain the precoding matrix at AP $i$, i.e., $\textbf{W}_{i}$, for each possible antenna subset.
During the iterative searches in segment $i$, the antenna subsets in other APs are fixed. 
The steps for generating the training samples are summarized in Algorithm 1, where $T$ is the total number of samples in the dataset of each CNN.
Overall, following the proposed IS algorithm, the size of the total search space reduces to $\tbinom{N}{M}I$, thereby greatly reducing the computational complexity. The numerical results in Section \ref{sec:simulation} show that the proposed IS algorithm can achieve a high sum spectral efficiency.

\begin{figure*}[!t]
\centerline{\includegraphics[width= 6.8in]{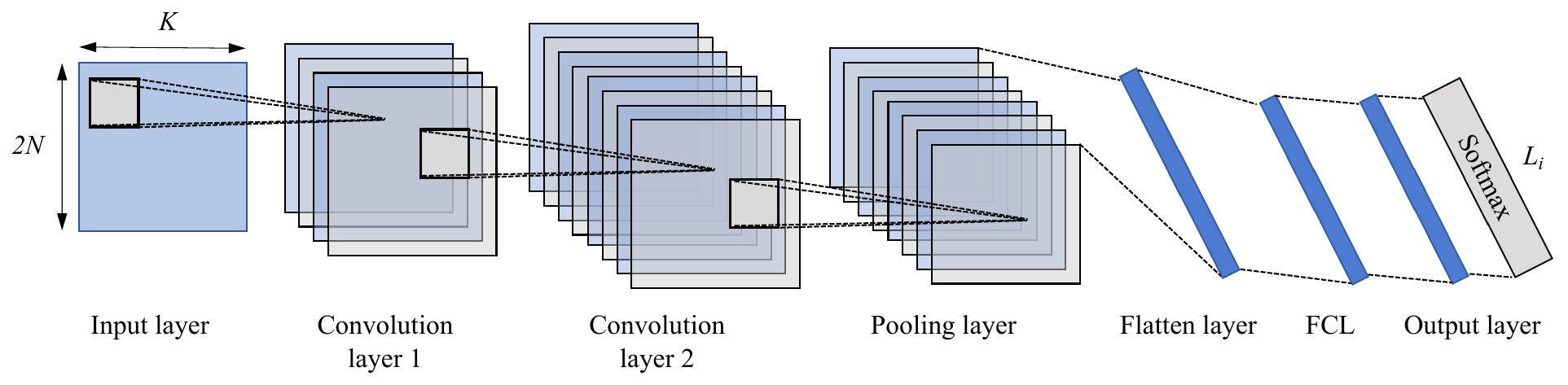}}
\caption{The structure of CNN $i$.}
\label{fig:CNN structure}
\end{figure*}

\subsection{Structure of the Proposed CNN Models}
We adopt CNN for antenna selection as it has a strong feature extraction ability and is good at processing  grid-structured data. In Section \ref{sec:CNN_Setup}, we will show that CNN-based antenna selection algorithm achieves a higher sum spectral efficiency than the GNN- and MLP-based algorithms.
We deploy a CNN at each AP that takes the locally estimated CSI as input and outputs the index of the optimal antenna subset.
The structure of the proposed CNN model at AP $i, \forall i \in \mathcal{I}$, is visualized in Fig. \ref{fig:CNN structure}. To be specific, the CNN model consists of an input layer, 2 convolution layers, a pooling layer, a flatten layer, a fully-connected layer (FCL) and an output layer. The input of CNN $i$ is the estimated local CSI at AP $i$, $\textbf{X}_{i}=[\Re\{\widehat{\textbf{H}}_{i}\}; \Im\{\widehat{\textbf{H}}_{i}\}]\in \mathbb{R}^{2N\times K}$. Followed by the input layer, 2 convolution layers are used to extract features from the input with the well-known ReLu, i.e., $f(x) = \text{max}(0, x)$, as activation function. A pooling layer is applied after the convolution layers to reduce the number of neural network parameters, improve the computational efficiency and reduce the risk of overfitting. The output of the pooling layer is then reshaped into a vector using a flattening layer. Subsequently, a FCL with ReLu as activation function is used between the flattening layer and the output layer to further extract information. Finally, the output layer consists of a ${\tbinom{N}{M}}\times 1$ FCL followed by a Softmax activation function given by
\begin{align}
p_{z_i^j} = \frac{e^{z_i^j}}{\sum_{j=1}^{J} e^{z_i^j}},
\end{align}
where $z_i^j$ denotes the $j$-th output of the FCL at AP $i$, $J = \tbinom{N}{M}$, and $p_{z_i^j}$ is the probability of choosing  the antenna subset with index $j$ at AP $i$. The output vector of the Softmax classifier is denoted by $\textbf{p}_{i}=[p_{z_i^1}, p_{z_i^2},\dots, p_{z_i^{J}}]$, and the index of the selected antenna subset is obtained as $L_i = \mathop{\arg\max}_{j} p_{z_i^j}$. Accordingly, $\mathcal{A}_{i}$  is set to the antenna set with index $L_i$. 
Note that modeling AS as a classification task may not be efficient in conventional massive MIMO systems where a BS may be equipped with a large number of antennas, and thus $J$ is an extremely large number. However, this is less a concern in cell-free MIMO systems where an AP typically has a small number of antennas.

\begin{figure*}[!t]
\centerline{\includegraphics[width= 6.3in]{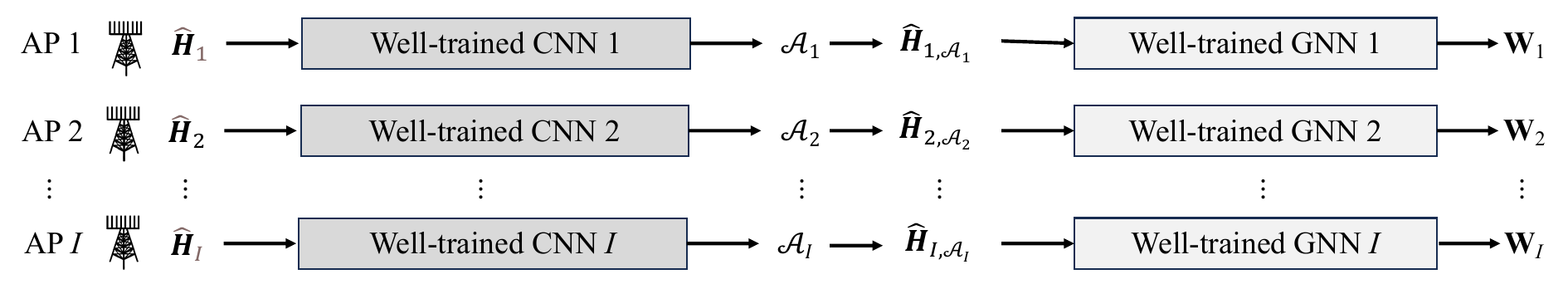}}
\caption{Online deployment of joint antenna selection and precoding.}
\label{fig:train_deploy_as}
\end{figure*}

\subsection{Offline Training of CNNs}
The offline training of CNNs can be performed at the CPU or the APs. If it is conducted at the APs, the training dataset needs to be downloaded to the APs. In contrast, if the training is performed at the CPU, the well-trained CNNs are downloaded to the APs for online inference. The latter can significantly reduce the fronthaul overhead. Each CNN is independently trained using its own training dataset.

In the presence of training dataset, supervised learning is adopted to train the CNNs. We choose the cross-entropy loss function, which is widely-used in classification models. To this end, the labels in the dataset are converted into one-hot vectors. More specifically, the label of sample $t$ in the dataset of CNN $i$, $L_i^t$, is converted into $ \textstyle \textbf{d}_i^t\in \{0,1\}^{\tbinom{N}{M}\times 1}$, where $[\textbf{d}_i^t]_{L_i^t}=1$ and $[\textbf{d}_i^t]_{l}=0, \forall l \ne L_i^t$. The cross-entropy loss function of CNN $i$
 is given by
\begin{align}
\mathcal{L}_{\mathrm{CNN},i} &= -\frac{1}{T}\sum_{t=1}^{T}\sum_{j=1}^{J} \left[\textbf{d}_i^t\right]_{j}\text{log}\left(p^t_{z_i^j}\right),
\end{align} 
where $T$ is the number of training samples and $p^t_{z_i^j}$ is the probability of choosing the antenna subset with index
$j$ at AP $i$ for the $t$-th training sample.

\subsection{Online Deployment of Joint Antenna Selection and Precoding}
The online deployment of joint antenna selection and precoding is performed in a totally distributed manner, as shown in Fig. \ref{fig:train_deploy_as}. Each AP downloads its own well-trained CNN and GNN from the CPU for local antenna selection and precoding design.
After AP $i$ obtains its locally estimated CSI, $\widehat{\textbf{H}}_{i}$, it selects antenna subset $\mathcal{A}_i$ using the well-trained CNN $i$. Subsequently, the estimated CSI of the selected antenna subset, $\widehat{\textbf{H}}_{i,\mathcal{A}_i}$, is taken as input to the well-trained GNN $i$. 
Finally, the precoding matrix at AP $i$, $\textbf{W}_i$, is obtained as the output of GNN $i$. Note that during the online deployment phase, both antenna selection and precoding do not require information exchange between APs.

The well-trained GNNs and CNNs only need to be updated when the characteristics of the wireless environment have experienced significant changes. Transfer learning can be used to improve learning efficiency during retraining.

\section{Performance Evaluation}
\label{sec:simulation}
This section presents the performance evaluation of our proposed joint CNN-based antenna selection and GNN-based precoding algorithm in comparison with the benchmark schemes.

\subsection{Simulation Setup}
\begin{figure}[!t]
\centerline{\includegraphics[width= 3.1in]{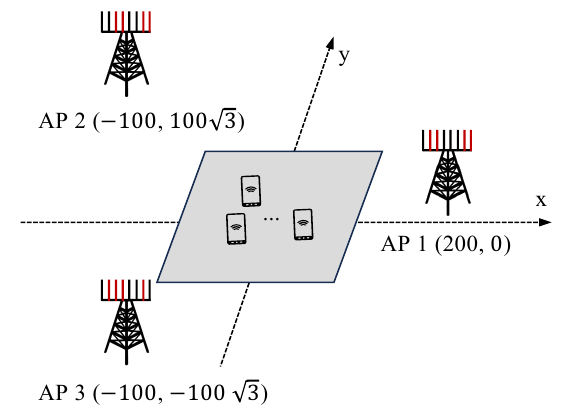}}
\caption{Simulation setup for the cell-free MIMO network  with $I=3$.}
\label{fig:simulation_setup}
\end{figure}

We consider a cell-free MIMO network consisting of multiple APs and users. 
The polar coordinate of AP $i$ is $(200\ \mathrm{m}, 2\pi (i-1)/I)$. The users are uniformly and randomly distributed within a square area  on the $xy$-plane with $x \in [-150,150]$ m and $y \in [-150,150]$ m.
A Cartesian coordinate illustration of $I=3$ is shown in Fig. \ref{fig:simulation_setup}.
 The large-scale path loss is given by 
\begin{align}
l\left ( d \right) &= l_{0} - 10 \alpha \textrm{log}_{10} \left ( {\frac {d} {d_{0}} } \right),
\end{align} 
where $l_{0} = -32.6$ dB represents the path loss at the reference distance, $d_{0}$, which is set to $1$ m, and $\alpha = 3.67$ represents the path loss exponent~\cite{Chen2023Secret}. 
Since we focus on downlink transmission, we have $\tau_{u}=0$. The noise power is computed by $\delta^{2}=-174\ \mathrm{dBm/Hz} + 10 \mathrm{log}10(W) + 7\ \mathrm{dB}$~\cite{Zaher2023}, where $W$ is the channel bandwidth.
The default simulation parameters are summarised in Table \ref{tab1} unless otherwise stated.

\begin{table}
\begin{center}
\caption{Default Simulation Parameters.}
\label{tab1}
\begin{tabular}{| c | c |}
\hline
\textbf{Parameters} &~~~\textbf{Value}~~~~\\
\hline
\hline
Channel bandwidth, $W$ & 20~MHz\\
%\hline
Noise power, $\delta^{2}$ & -94~dBm\\
%\hline
Coherence block length, $\tau_{c}$ & 200\\
Pilot signal length, $\tau_{p}$ & 10\\
Number of APs, $I$ & 3\\
%\hline
Maximum transmit power per AP, $P_{\textrm{max}}$ & $20$ dBm\\
%\hline
Number of users, $K$ & 4\\
%\hline
Number of antennas per AP, $N$ & 8\\
%\hline 
Number of active antennas per AP, $M$ & 5\\
\hline 
\end{tabular}
\end{center}
\end{table}

\subsection{GNN Setup}
\label{sec:GNN_Setup}
In this subsection, we present the network parameters of GNN $i$, $\forall i \in \mathcal{I}$, the structure of which has been shown in Fig. \ref{fig:GNN i}. We set the number of graph convolution layers to $L=2$. 
In each graph convolution layer as shown in Fig. \ref{fig:graph convolution}, the two MLPs,  $\alpha_{i}^{l} (\cdot)$ and $\delta_{i}^{l} (\cdot)$, $\forall i\in \mathcal{I}, l\in \mathcal{L}$, share the same network architecture; each MLP has two layers with 800 and 400 neurons, respectively.
 Leaky ReLU is employed as the activation function, which is given by
\begin{align}
g(x) = \begin{cases}
  x, & \text{if } x \geq 0 \\
  \mu x, & \text{otherwise},
\end{cases}
\end{align}
where $\mu = 0.1$ is a slope to avoid the gradient being zero when the input is less than zero. 

\begin{figure}[!t]
\centerline{\includegraphics[width= 3.6in]{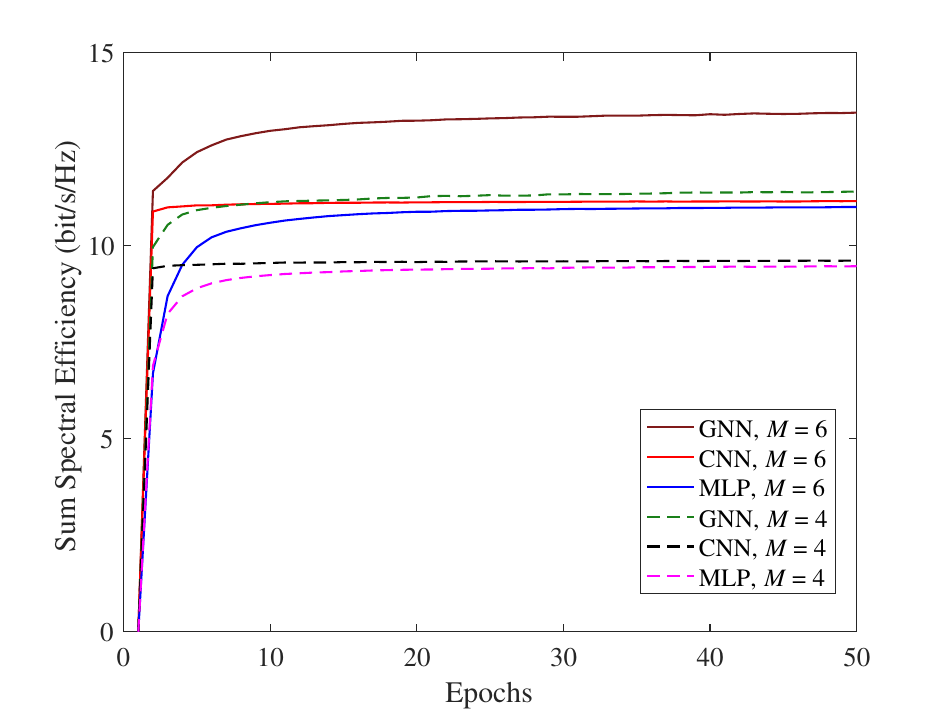}}
\caption{Sum spectral efficiency versus training epochs for precoding design.}
\label{fig:converge}
\end{figure}

During the training stage, the GNNs are optimized with Adam optimizer using stochastic gradient descent (SGD). 
The training process lasts for 100 epochs. In each epoch, the model undergoes 100 iterations, where 600 groups of random user locations are sampled in each iteration. 
The initial learning rate is set to 0.001 and is decreased by a factor of 0.995 after every 100 iterations.
  The convergences of the proposed distributed GNN-based algorithm is shown in Fig. \ref{fig:converge}. We can see that the algorithm converges quickly under various system parameters. In particular, a satisfactory sum spectral efficiency can be achieved after 10 training epochs.
  In addition, we have developed CNN- and MLP-based baseline algorithms to perform the precoding design as comparisons to the proposed GNN-based algorithm. The network parameters have been fine-tuned to achieve the best performance.  We see that the GNN-based algorithm outperforms the two baseline algorithms in terms of the sum spectral efficiency.

\subsection{CNN Setup}
\label{sec:CNN_Setup}
The structure of CNN $i$, $\forall i \in \mathcal{I}$, has been shown in Fig. \ref{fig:CNN structure}.
The first convolution layer has 50 different $3\times 2$ filters and the second convolution layer has 50 different $3\times 1$ filters. A max pooling layer with a $2\times 2$ filter is employed to down-sample the feature maps outputed by the convolution layers.
The FCL between the flatten layer and the output layer has 128 neurons. The CNNs are trained for 50 epochs using the Adam optimizer with a batch size of 64.

\begin{figure}[!t]
\centerline{\includegraphics[width= 3.6in]{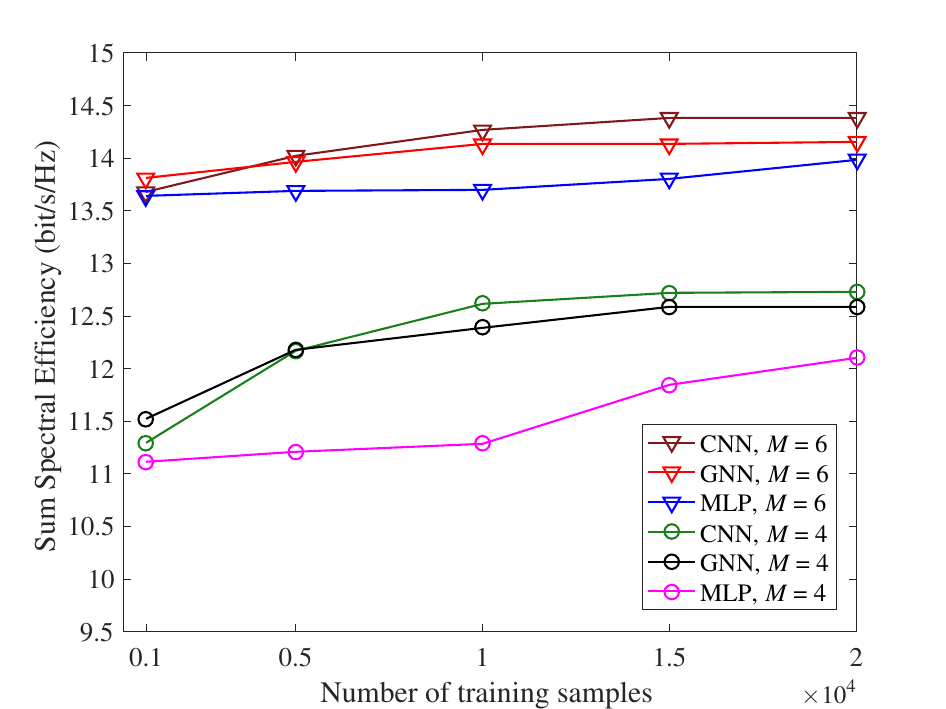}}
\caption{Sum spectral efficiency versus the size of training set for antenna selection.}
\label{fig:training_size}
\end{figure}

 Fig. \ref{fig:training_size} illustrates the impact of the number of training samples per dataset on the sum spectral efficiency.  We compare the proposed CNN-based antenna selection algorithm with the GNN- and MLP- based baseline algorithms.
 As can be observed, the CNN-based antenna selection algorithm achieves a higher spectral efficiency than the baseline algorithms.
 Moreover, it is observed that for the proposed CNN-based algorithm, the sum spectral efficiency grows rapidly as the number of training samples increases from $10^{3}$ to $10^{4}$. Then the growth trend slows down and the sum spectral efficiency saturates when the number of training samples reaches $3 \times 10^{4}$.
 As such, we choose $3 \times 10^{4}$ as the size of the training dataset for the training of each CNN.

\subsection{Performance Comparison}
This subsection demonstrates the superiority of the proposed joint CNN-based antenna selection and GNN-based precoding algorithm in comparison with the existing benchmark schemes. 
All the schemes are described as follows:
\begin{itemize}

\item Centralized MMSE + IS: Centralized MMSE
precoding~\cite{bjornson2019making} is computed at the CPU using the global estimated CSI. 
The heuristic power allocation from~\cite{interdonato2019scalability} is adopted, where more power is allocated to users with better channel conditions. Specifically, the power that AP $i$ allocates to user $k$ is $\textstyle P_{i,k}=P_{\mathrm{max}}\frac{|\widehat{\textbf{h}}_{i,k,\mathcal{A}_{i}}|}{\sum_{j\in \mathcal{K}}|\widehat{\textbf{h}}_{i,j,\mathcal{A}_{i}}|}$.
The proposed IS algorithm in Section \ref{sec:dataset} is adopted for antenna selection.

\item Centralized MMSE: The centralized MMSE precoding matrix is computed at the CPU by collecting the global estimated CSI. The heuristic power allocation and the random antenna selection strategy are adopted.

\item Distributed MMSE: Each AP computes its local MMSE precoding matrix using its local estimated CSI. The heuristic power allocation and the random antenna selection strategy are adopted.

\item MRT: MRT precoding~\cite{7827017} focuses on maximizing the power of the desired signals. The heuristic power allocation and the random antenna selection strategy are adopted. The precoding matrix at each AP is the same as its locally estimated channel matrix, and thus no CSI exchange is required.

\item GNN: The proposed distributed GNN-based precoding scheme with a random antenna selection strategy.

\item GNN + CNN: The proposed distributed joint CNN-based antenna selection and GNN-based precoding algorithm.

\end{itemize}
Although there exist more advanced centralized power allocation algorithms~\cite{7827017, nayebi2017precoding}, they are computationally extensive. Hence, we use the heuristic power allocation in the benchmark schemes, which is known to work fairly well.

\begin{figure}[!t]
\centerline{\includegraphics[width= 3.6in]{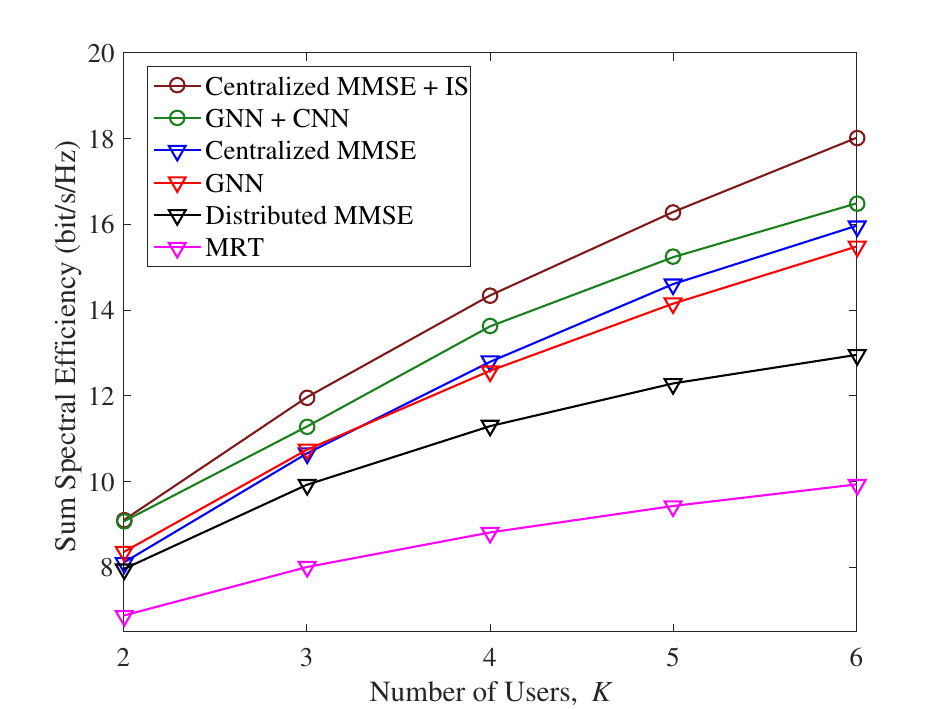}}
\caption{Sum spectral efficiency versus the number of users.}
\label{fig:user}
\end{figure}

In Fig. \ref{fig:user}, we show the achievable sum spectral efficiency versus different numbers of users. We can see that the centralized MMSE + IS scheme achieves the highest sum spectral efficiency among the considered schemes. However, it requires global CSI and has a high computational complexity. Notably, the proposed GNN + CNN scheme significantly outperforms the distributed benchmark schemes and slightly outperforms the centralized MMSE scheme for various numbers of users. This indicates that the proposed GNN + CNN scheme can achieve a sum spectral efficiency comparable to the centralized  precoding scheme.
In addition, we see that the centralized MMSE + IS and GNN + CNN schemes significantly outperform their counterparts without antenna selection, indicating the importance of antenna selection in improving the sum spectral efficiency.
The MRT scheme achieves the lowest sum spectral efficiency among all considered schemes, as it fails to suppress the inter-user interference.

\begin{figure}[!t]
\centerline{\includegraphics[width= 3.6in]{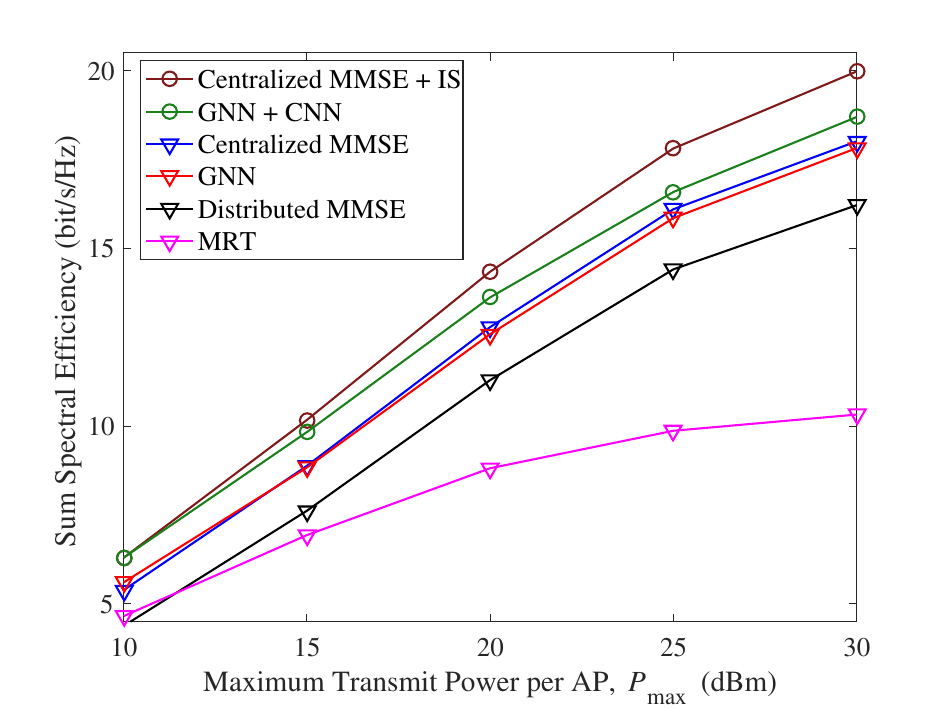}}
\caption{Sum spectral efficiency versus the maximum transmit power.}
\label{fig:BSpower}
\end{figure}

Fig. \ref{fig:BSpower} illustrates the sum spectral efficiency versus the maximum transmit power per AP. 
It is straightforward to see that the
sum spectral efficiency monotonically increases with the maximum transmit power per AP.  When the APs have small transmit power budgets, the proposed GNN + CNN scheme achieves a sum spectral efficiency close to the centralized MMSE + IS scheme. 
It is also observed that the proposed GNN scheme outperforms the distributed MMSE scheme in terms of the sum spectral efficiency.

\begin{figure}[!t]
\centerline{\includegraphics[width= 3.6in]{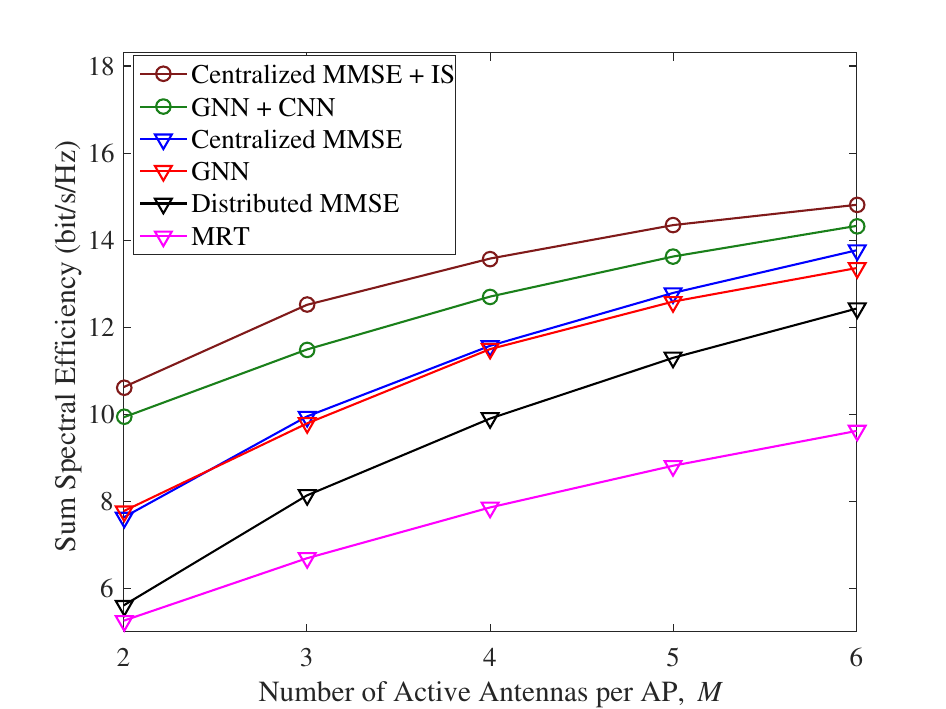}}
\caption{Sum spectral efficiency versus the number of active antennas.}
\label{fig:active_antenna}
\end{figure}

In Fig. \ref{fig:active_antenna}, we present the sum spectral efficiency versus the number of active antennas per AP. We observe that for a given number of antennas, a higher sum spectral efficiency is achieved when
more antennas are activated.
Generally, the gain of antenna selection in terms of sum spectral efficiency is more significant when fewer antennas are activated.
This demonstrates the importance of antenna selection for a small number of active antennas per AP.

\begin{figure}[!t]
\centerline{\includegraphics[width= 3.6in]{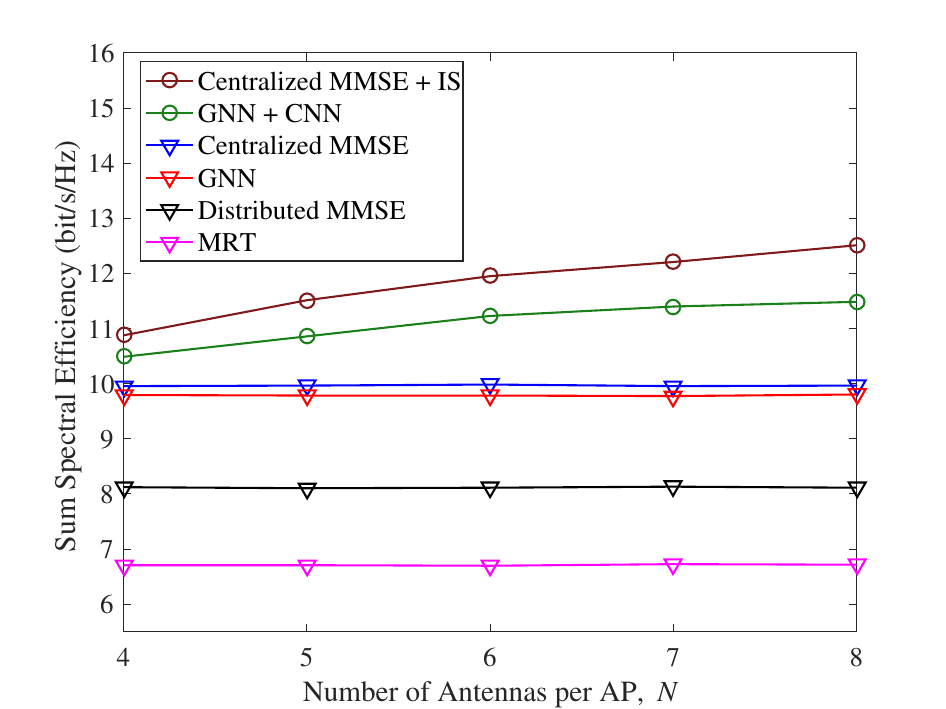}}
\caption{Sum spectral efficiency versus the number of antennas per AP with $M=3$.}
\label{fig:all_antenna}
\end{figure}

In Fig. \ref{fig:all_antenna}, we show the impact of the number of antennas per AP on the sum spectral efficiency. 
In particular, the centralized MMSE,  distributed MMSE,  MRT and GNN schemes adopt the random antenna selection strategy, and thus the sum spectral efficiency of these schemes remains unchanged with the number of antennas per AP.
In contrast, for a given number of active antennas per AP, the GNN + CNN and centralized MMSE + IS schemes can achieve a higher sum spectral efficiency when the number of antennas per AP increases. This indicates the necessity of antenna selection when each AP is equipped with a large number of antennas. 

\begin{figure}[!t]
\centerline{\includegraphics[width= 3.6in]{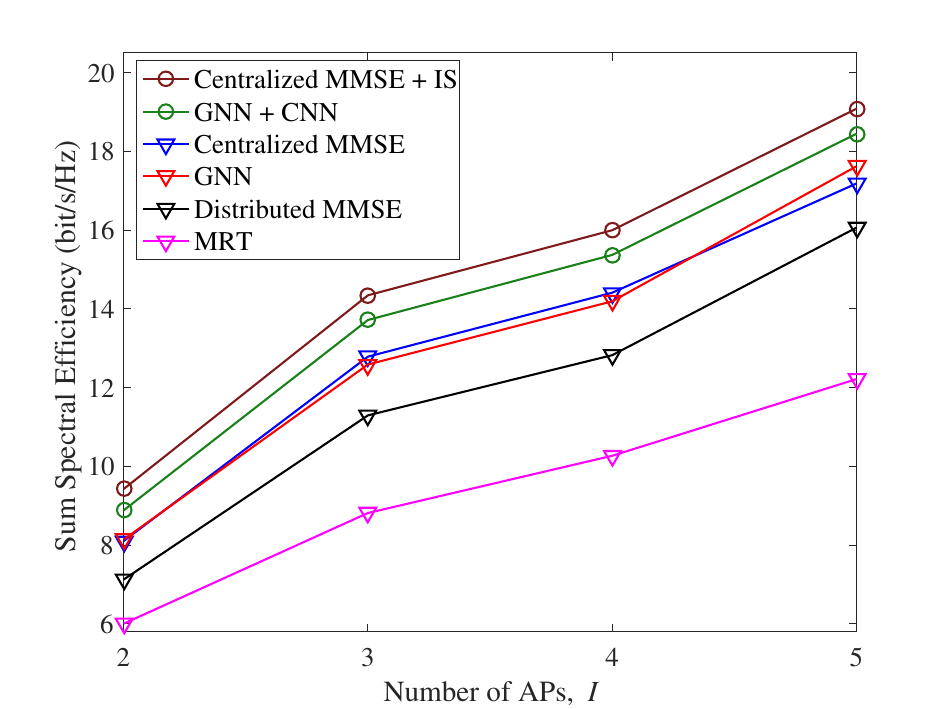}}
\caption{Sum spectral efficiency versus the number of APs.}
\label{fig:bs}
\end{figure}

Fig. \ref{fig:bs} shows the impact of the number of APs on the sum spectral efficiency. 
We see that the proposed GNN + CNN scheme scales well with the number of APs. In particular, the proposed GNN + CNN scheme outperforms the distributed MMSE scheme by a large margin.
It also achieves a higher sum spectral efficiency than the centralized MMSE scheme for different number of APs.

Overall, we can conclude from Fig. \ref{fig:user}-\ref{fig:bs} that the proposed distributed GNN-based precoding scheme is effective in achieving a high sum spectral efficiency. 
In addition, the proposed fully distributed joint GNN-based precoding and CNN-based antenna selection scheme achieves a sum spectral efficiency comparable to its centralized counterpart for various system parameters.
 
In Table \ref{tab2}, we present the sum spectral efficiency, computational time, and information exchange of the proposed GNN + CNN scheme and the centralized MMSE + IS scheme. The sum spectral efficiency and computational time are averaged over 1000 independent channel realizations.  We implement all the considered schemes using Python 3.8 on a laptop with a 4 core Intel Core i7 CPU with 2.2 GHz base frequency and 16 GB memory. The information exchange consists of CSI transmitted from the APs to the CPU and precoding vectors conveyed from the CPU to the APs in a coherence block. 
We see that compared to the centralized MMSE + IS scheme, the proposed GNN + CNN scheme achieves $95.6\%$ sum spectral efficiency with only $0.33\%$ computational time. Moreover, the proposed GNN + CNN scheme is fully distributed, requiring no information exchange. Therefore, the proposed GNN + CNN scheme can achieve real-time processing and alleviate the fronthaul overhead. Although the computational complexity of the proposed GNN + CNN scheme is slightly higher than that of the baseline schemes without antenna selection, the proposed GNN + CNN scheme provides a significant sum spectral efficiency gain.

\begin{table*}
\begin{center}
\caption{Comparison of Sum Spectral Efficiency, Computational Time, and Information Exchange.}
\label{tab2}
\begin{tabular}{| c | c | c | c |}
\hline
 \textbf{Algorithm} & \textbf{Sum spectral efficiency (bit/s/Hz)} & \textbf{Computational time (ms)} &  \textbf{Information exchange (complex scalars)} \\
\hline
\hline
Centralized MMSE + IS & 15.1 & 227.6 & $INK + IMK$  \\
\hline
GNN + CNN & 14.44 & 0.76 & 0 \\
\hline
Centralized MMSE & 11.89 & 0.43 & $2IMK$  \\
\hline
Distributed MMSE & 13.47 & 0.37 & $0$  \\
\hline
\end{tabular}
\end{center}
\end{table*}

\section{Conclusion}
\label{sec:conclusion}
In this paper, the joint design of antenna selection and precoding in the downlink of a cell-free MIMO network has been addressed. 
We have formulated a sum spectral efficiency maximum problem, considering realistic pilot-aided CSI acquisition.
A novel fully distributed machine learning algorithm has been proposed to maximize the sum spectral efficiency. 
More specifically, each AP deploys a CNN for antenna selection and a GNN for precoding design using only locally estimated CSI as input, and so alleviating the fronthaul overhead. 
Unsupervised learning has been used to train the GNNs, bypassing the time-consuming phase of dataset generation.
The proposed distributed machine learning algorithm effectively finds the  mapping relationship between the local domain input and the desired antenna selection as well as precoding design.  
Simulation results illustrate that the proposed distributed machine learning algorithm achieves a sum spectral efficiency close to its centralized counterpart with a much less computational time. In the future, it will be of interest to investigate joint antenna selection and user association in a user-centric cell-free MIMO network.

\bibliographystyle{IEEEtran}
\bibliography{IEEEabrv, mainbib}

\end{document}